\documentclass[11pt]{elsart}
\usepackage[dvips]{graphicx}

\begin{document}
\begin{frontmatter}

\title{A general geometric growth model for pseudofractal scale-free web}

\author{Zhongzhi Zhang\corauthref{zz}}
\corauth[zz]{Corresponding author.} \ead{zhangzz@fudan.edu.cn}
\address{Department of Computer Science and Engineering, Fudan
University, Shanghai 200433, China}%
\address{Shanghai Key Lab of Intelligent Information Processing,
Fudan University, Shanghai 200433, China}

\author{Lili Rong}
\ead{llrong@dlut.edu.cn}
\address{
Institute of Systems Engineering, Dalian University of Technology, Dalian 116023, Liaoning, China}%

\author{Shuigeng Zhou}
\ead{sgzhou@fudan.edu.cn}
\address{Department of Computer Science and Engineering, Fudan
University, Shanghai 200433, China}%
\address{Shanghai Key Lab of Intelligent Information Processing,
Fudan University, Shanghai 200433, China}

\begin{abstract}
We propose a general geometric growth model for pseudofractal
scale-free web, which is controlled by two tunable parameters. We
derive exactly the main characteristics of the networks: degree
distribution, second moment of degree distribution, degree
correlations, distribution of clustering coefficient, as well as the
diameter, which are partially determined by the parameters.
Analytical results show that the resulting networks are
disassortative and follow power-law degree distributions, with a
more general degree exponent tuned from 2 to $1+\frac{\ln3}{\ln2}$;
the clustering coefficient of each individual node is inversely
proportional to its degree and the average clustering coefficient of
all nodes approaches to a large nonzero value in the infinite
network order; the diameter grows logarithmically with the number of
network nodes. All these reveal that the networks described by our
model have small-world effect and scale-free topology.

%\\PACS: 02.50.Cw, 05.45Pq, 89.75.-k, 05.10-a

\end{abstract}

%89.20.Hh World Wide Web, Internet
%89.75.Da Systems obeying scaling laws
%89.75.Fb Structures and organization in complex systems
%89.75.-k Complex systems
%89.75.Hc Networks and genealogical trees

\date{}
\end{frontmatter}

%%%%%%%%%%%%%%%%%%%%%%%%%%%%%%%%%%%%%%%%%%%%%%%%%%%%%%%%%%%%%%%%%
%%%%%%%%%%%%%%%%%%%%%%%%%%%%%%%%%%%%%%%%%%%%%%%%%%%%%%%%%%%%%%%%%
%\vskip -0.5cm\color{Blue}
%\vbox to 0pt{\kern -14cm {
%\noindent \small \copyright 2005
%{\em Elsevier Science B.V. All rights reserved}\\
%{\em Physica A}, submitted.}
%\vss}\color{Black}

%%%%%%%%%%%%%%%%%%%%%%%%%%%%%%%%%%%%%%%%%%%%%%%%%%%%%%%%%%%%%%%%%%%%
\section{Introduction}
Since the pioneering papers by Watts and Strogatz on small-world
networks \cite{WaSt98} and Barab\'asi and Albert on scale-free
networks \cite{BaAl99}, complex networks, which describe many
systems in nature and society, have become an area of tremendous
recent interest \cite{St01,AlBa02,DoMe02,Ne03,BoLaMo06}. In the last
few years, modeling real-life systems has attracted an exceptional
amount of attention within the physics community. While a lot of
models have been proposed, most of them are stochastic
\cite{St01,AlBa02,DoMe02,Ne03,BoLaMo06}. However, because of their
advantages, deterministic networks have also received much
attention~\cite{BaRaVi01,IgYa05,DoGoMe02,CoFeRa04,JuKiKa02,RaSoMoOlBa02,RaBa03,No03,NaUeKaAk05,AnHeAnSi05,DoMa05,RoKiBobe05,Bobe05,HiBe06,ZhCoFeRo05,ZhRo05,CoOzPe00,CoSa02,ZhRoGo05,ZhRoCo05a,ZhWaHuCh04,co04,Ac04,ChDa05}.
First, the method of generating deterministic networks makes it
easier to gain a visual understanding of how networks are shaped,
and how do different nodes relate to each other~\cite{BaRaVi01};
moreover, deterministic networks allow to compute analytically their
properties: degree distribution, clustering coefficient, average
path length, diameter, betweenness, modularity and adjacency matrix
whose eigenvalue spectrum characterizes the
topology~\cite{BaRaVi01,IgYa05,DoGoMe02,CoFeRa04,JuKiKa02,RaSoMoOlBa02,RaBa03,No03,NaUeKaAk05,AnHeAnSi05,DoMa05,RoKiBobe05,Bobe05,HiBe06,ZhCoFeRo05,ZhRo05,CoOzPe00,CoSa02,ZhRoGo05,ZhRoCo05a,ZhWaHuCh04,co04,Ac04,ChDa05}.

The first model for deterministic scale-free networks was proposed
by Barab\'asi \emph{et al.} in Ref.~\cite{BaRaVi01} and was
intensively studied in Ref.~\cite{IgYa05}. Another elegant model,
called pseudofractal scale-free web (PSW)~\cite{DoGoMe02}, was
introduced by Dorogovtsev, Goltsev, and Mendes, and was extended by
Comellas \emph{et al.} in Ref.~\cite{CoFeRa04}. Based on a similar
idea of PSW, Jung \emph{et al.} presented a class of recursive
trees~\cite{JuKiKa02}. Additionally, in order to discuss modularity,
Ravasz \emph{et al.} proposed a hierarchical network
model~\cite{RaSoMoOlBa02,RaBa03}, the exact scaling properties and
extensive study of which were reported in Refs.~\cite{No03}
and~\cite{NaUeKaAk05}, respectively. Recently, In relation to the
problem of Apollonian space-filing packing, Andrade \emph{et al.}
introduced Apollonian networks~\cite{AnHeAnSi05} which were also
proposed by Doye and Massen in Ref.~\cite{DoMa05} and have been
intensively investigated
\cite{ZhCoFeRo05,ZhRo05,ZhYaWa05,ZhRoCo05b,AnHe05,LiGaHe04,AnMi05}.
In addition to the above models, deterministic networks can be
created by various techniques: modification of some regular
graphs~\cite{CoOzPe00}, addition and product of
graphs~\cite{CoSa02}, edge iterations~\cite{ZhRoGo05,ZhRoCo05a} and
other mathematical methods as in
Refs.~\cite{ZhWaHuCh04,co04,Ac04,ChDa05}.

As mentioned by Barab\'asi \emph{et al.}, it would be of major
theoretical interest to construct deterministic models that lead to
scale-free networks~\cite{BaRaVi01}. Here we do an extensive study
on pseudofractal scale-free web~\cite{DoGoMe02}. The PSW can be
considered as a process of edge multiplication. In fact, a clique
(edge is a special case of it) can also reproduce new cliques and
the number of the new reproduction may be different at a time.
Motivated by this, in a simple recursive way we propose a general
model for PSW by including two parameters, with PSW as a particular
case of the present model. The deterministic construction of our
model enables one to obtain the analytic solutions for its structure
properties. By adjusting the parameters, we can obtain a variety of
scale-free networks.

\section{The general geometric growth model}

Before introducing our model we give the following definitions on a
graph (network). The term \emph{size} refers to the number of edges
in a graph. The number of nodes in a graph is called its
\emph{order}. When two nodes of a graph are connected by an edge,
these nodes are said to be \emph{adjacent}, and the edge is said to
join them. A \emph{complete graph} is a graph in which all nodes are
adjacent to one another. Thus, in a complete graph, every possible
edge is present. The complete graph with $q$ nodes is denoted as
$K_q$ (also referred in the literature as $q$-\emph{clique}). Two
graphs are \emph{isomorphic} when the nodes of one can be relabeled
to match the nodes of the other in a way that preserves adjacency.
So all $q$-cliques are isomorphic to one another.

%%%%%%%%%%%%%%%%%%%%%%%%%%%%%%%%%%%%%%%%%%%%%%%%%%%%%%%%%%
% Figure  1
%%%%%%%%%%%%%%%%%%%%%%%%%%%%%%%%%%%%%%%%%%%%%%%%%%%%%%%%%%
\begin{figure}
\begin{center}
\includegraphics[width=13cm]{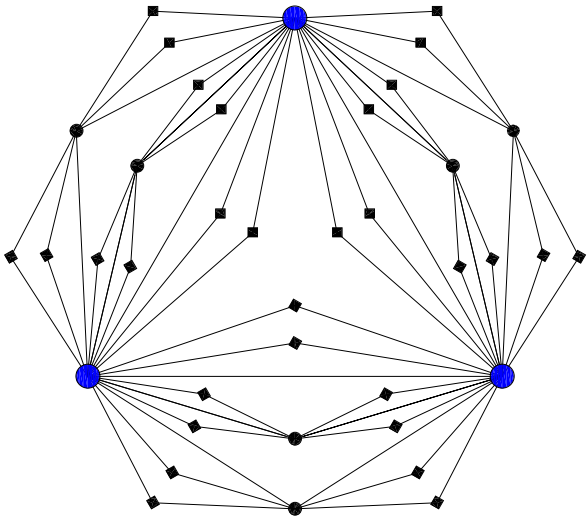}
\caption{Scheme of the growth of the deterministic network for the
case of $m=2$ and $q=2$. Only the first three steps are shown.}
\label{recursive}
\end{center}
\end{figure}
%%%%%%%%%%%%%%%%%%%%%%%%%%%%%%%%%%%%%%%%%%%%%%%%%%%%%%%%%%

The network is constructed in a recursive way. We denote the network
after $t$ steps by $R(q,t)$, $q\geq 2, t\geq 0$ (see Fig.
\ref{recursive}). Then the network at step $t$ is constructed as
follows: For $t=0$, $R(q,0)$ is a complete graph $K_{q+1}$ (or
$(q+1)$-clique) consist of $q+1$ $q$-cliques), and $R(q,0)$ has
$q+1$ nodes and $q(q+1)/2$ edges. For $t\geq 1$, $R(q,t)$ is
obtained from $R(q,t-1)$ by adding $m$ new nodes for each of its
existing subgraphs isomorphic to a $q$-clique, and each new node is
connected to all the nodes of this subgraph. In the special case
$m=1$ and  $q=2$, it is reduced to the pseudofractal scale-free web
described in Ref. \cite{DoGoMe02}. In the limiting case of $m=1$, we
obtain the same networks as in Ref.~\cite{CoFeRa04}. However, our
family is richer as $m$ can take any natural value.

\begin{figure}
\begin{center}
\includegraphics[width=12cm]{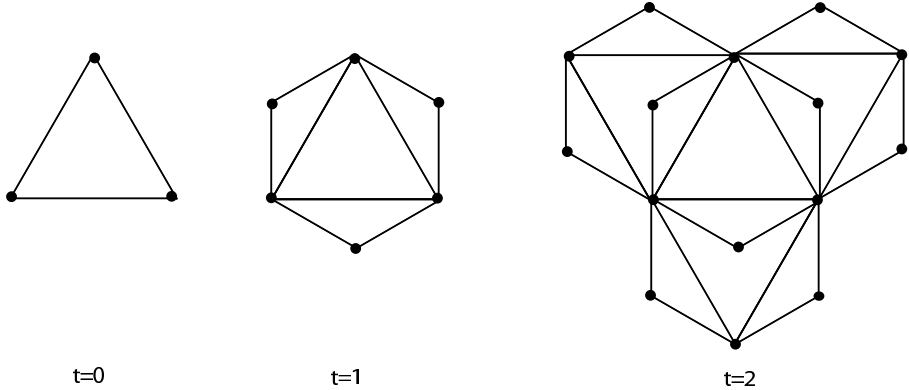}
\caption{Growth process of pseudofractal scale-free web (i.e. the
case of $m=1$ and $q=2$), exhibiting the first three steps.}
\label{pseudofractal}
\end{center}
\end{figure}

\begin{figure}
\begin{center}
\includegraphics[width=12cm]{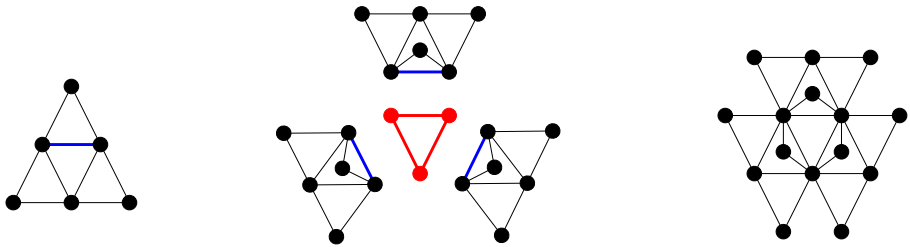}
\caption{Aggregation process from $R(2,1)$ to $R(2,2)$, which is
obtained by adjoining of three copies of $R(2,1)$ at the hubs.}
\label{aggregation}
\end{center}
\end{figure}

There is an interpretation called `aggregation' \cite{GrKa82} for
our model. As an example, here we only explain them for the case of
$m=1$ and $q=2$. Figure \ref{pseudofractal} illustrates the growing
process for this particular case, which may be accounted for as an
`aggregation' process described in detail as follows. First, three
of the initial triangle ($t=0$) are assembled to form a new unit
($t=1$). Then we assemble three of these units at the hubs (the
nodes with highest degree) in precise analogy with the step leading
from $t=0$ to $t=1$ to form a new cell ($t=2$) (see Fig.
\ref{aggregation}). This process can be iterated an arbitrary number
of times. Moreover, an alternative explanation of our model which is
often useful is that of `miniaturization' (see Ref. \cite{GrKa82}).

%%%%%%%%%%%%%%%%%%%%%%%%%%%%%%%%%%%%%%%%%%%%%%%%%%%%%%%%%%
% Figure  2
%%%%%%%%%%%%%%%%%%%%%%%%%%%%%%%%%%%%%%%%%%%%%%%%%%%%%%%%%%
%%%%%%%%%%%%%%%%%%%%%%%%%%%%%%%%%%%%%%%%%%%%%%%%%%%%%%%%%%

%%%%%%%%%%%%%%%%%%%%%%%%%%%%%%%%%%%%%%%%%%%%%%%%%%%%%%%%%%
% Figure  3
%%%%%%%%%%%%%%%%%%%%%%%%%%%%%%%%%%%%%%%%%%%%%%%%%%%%%%%%%%
%%%%%%%%%%%%%%%%%%%%%%%%%%%%%%%%%%%%%%%%%%%%%%%%%%%%%%%%%%

%%%%%%%%%%%%%%%%%%%%%%%%%%%%%%%%%%%%%%%%%%%%%%%%%%%%%%%%%%%%%%%%
\section{Structural properties}

Below we will find that the tunable parameters $m$ and $q$ control
some relevant characteristics of the network $R(q,t)$. Because $q=2$
is a particular case, for conveniences, we treat $q=2$ and $q\geq3$
separately.

\subsection{Limiting case of $q=2$}

\emph{Order and size.} In the case of $q=2$, we denote $R(q,t)$ by
$\mathcal{G}(t)$. Let us consider the total number of nodes $N_t$
and total number of edges $E_t$ in $\mathcal{G}(t)$. Denote $n_v(t)$
as the number of nodes created at step $t$. Note that the addition
of each new node leads to two new edges. By construction, for $t\geq
1$, we have
\begin{equation}
n_v(t)=mE_{t-1}
\end{equation}
and
\begin{equation}
E_t=E_{t-1}+2n_v(t).
\end{equation}
Considering the initial condition $n_v(0)=3$ and $E_0=3$, it follows
that
\begin{equation}\label{Nv1}
n_v(t)=3m\,(2m+1)^{t-1} \quad\ \hbox{and}\quad\ E_t=3\,(2m+1)^{t}.
\end{equation}
Then the number of nodes increases with time exponentially and the
total number of nodes present at step $t$ is
\begin{eqnarray}\label{Nt1}
N_t=\sum_{t_i=0}^{t}n_v(t_i)=\frac{3\,(2m+1)^{t}+3}{2}.
\end{eqnarray}
Thus for large $t$, The average degree $\overline{k}_t=
\frac{2E_t}{N_t}$ is approximately $4$.

%%%%%%%%%%%%%%%%%%%%%%%%%%%%%%%%%%%%%%%%%%%%%%%%%%%%%%%%%%%%%%%%

%\subsection{Degree distribution}

\emph{Degree distribution.} Let $k_i(t)$ be the degree of node $i$
at step $t$. Then by construction, it is not difficult to find
following relation:
\begin{equation}
k_i(t)=(m+1)k_i(t-1),
\end{equation}
which expresses a preference attachment \cite{BaAl99}. If node $i$
is added to the network at step $t_i$, $k_i(t_i)=2$ and hence
\begin{equation}
k_i(t)=2\,(m+1)^{t-t_i}\label{Ki1}.
\end{equation}
Therefore, the degree spectrum of the network is discrete. It
follows that the degree distribution is given by
\begin{equation}
P(k)=\left\{\begin{array}{lc} {\displaystyle{n_v(0)\over
N_t}={2\over (2m+1)^{t}+1} }
& \ \hbox{for}\ t_i=0\\
{\displaystyle{n_v(t_i)\over N_t}={2m\,(2m+1)^{t_i-1}\over
 (2m+1)^{t}+1} }
& \ \hbox{for}\  t_i\ge 1\\
0 & \ \hbox{otherwise}\end{array} \right.
\end{equation}
and that the cumulative degree distribution \cite{Ne03,DoGoMe02} is
\begin{equation}
P_{\rm cum}(k)=\sum_{\rho \leq t_i}\frac{n_v(\rho)}{N_t}
={(2m+1)^{t_i}+1\over (2m+1)^{t}+1}.
\end{equation}
Substituting for $t_i$ in this expression using
$t_i=t-\frac{\ln(k/2)}{\ln (m+1)}$ gives
\begin{eqnarray}\label{gamma1}
P_{\rm cum}(k)&=&{(2m+1)^{t}\left({k/2}\right)^{-\frac{\ln(2m+1)}{\ln(m+1)}}+1\over (2m+1)^{t}+1}\nonumber\\
          &\approx&\left({k\over2}\right)^{\textstyle-{\ln(2m+1)\over \ln(m+1)}}\qquad\hbox{for large $t$}.
\end{eqnarray}
So the degree distribution follows the power law with the exponent
$\gamma=1+{\frac{\ln(2m+1)}{\ln(m+1)}}$. For the particular case of
$m=1$, Eq. (\ref{gamma1}) recovers the result previously obtained in
Ref. \cite{DoGoMe02}.

%\subsection{Second moment of degree distribution}

\emph{Second moment of degree distribution.} Let us calculate the
second moment of degree distribution $\langle k^{2}\rangle$. It is
defined by
\begin{eqnarray}\label{Ki21}
 \langle k^{2}\rangle =\frac{1}{N_t}\sum_{t_i=0}^{t}n_v(t_i)\left[k(t_i,t)\right]^{2},
\end{eqnarray}
where $k(t_i,t)$ is the degree of a node at step $t$, which was
generated at step $t_i$. This quality expresses the average of
degree square over all nodes in the network. It has large impact on
the dynamics of spreading \cite{PaVe01a,PaVe01b} and the onset of
percolation transitions \cite{AlJeBa00,CoerAvSa01} taking place in
networks. when $\langle k^{2}\rangle$ is diverging, the networks
allow the onset of large epidemics whatever the spreading rate of
the infection \cite{PaVe01a,PaVe01b}, at the same time the networks
are extremely robust to random damages, in other words, the
percolation transition is absent \cite{AlJeBa00,CoerAvSa01}.

substituting Eqs. (\ref{Nv1}), (\ref{Nt1}) and (\ref{Ki1}) into Eq.
(\ref{Ki21}), we derive
\begin{eqnarray}\label{Ki22}
\langle k^{2} \rangle
&=&\frac{8}{1+\frac{1}{(2m+1)^{t}}}\left[\left(\frac{1}{m}+1
\right)\left(\frac{(m+1)^{2}}{2m+1}\right)^{t}-
\frac{1}{m}\right]\nonumber\\
&\approx&\frac{8\,(m+1)^{2t+1}}{m(2m+1)}\rightarrow\infty
\qquad\hbox{for large $t$.}
\end{eqnarray}
In this way, second moment of degree distribution $\langle
k^{2}\rangle$ has been calculated explicitly, and result shows that
it diverges as an exponential law. So the networks are resilient to
random damage and are simultaneously sensitive to the spread of
infections.

%\subsection{Degree correlations}

\emph{Degree correlations.} As the field has progressed, degree
correlation \cite{MsSn02,PaVaVe01,VapaVe02,Newman02,Newman03c} has
been the subject of particular interest, because it can give rise to
some interesting network structure effects. An interesting quantity
related to degree correlations is the average degree of the nearest
neighbors for nodes with degree $k$, denoted as $k_{\rm nn}(k)$
\cite{PaVaVe01,VapaVe02}. When $k_{\rm nn}(k)$ increases with $k$,
it means that nodes have a tendency to connect to nodes with a
similar or larger degree. In this case the network is defined as
assortative \cite{Newman02,Newman03c}. In contrast, if $k_{\rm
nn}(k)$ is decreasing with $k$, which implies that nodes of large
degree are likely to have near neighbors with small degree, then the
network is said to be disassortative. If correlations are absent,
$k_{\rm nn}(k)=const$.

We can exactly calculate $k_{\rm nn}$ for the networks using Eq.
(\ref{Ki1}) to work out how many links are made at a particular step
to nodes with a particular degree. Except for three initial nodes
generated at step 0, no nodes born in the same step, which have the
same degree, will be linked to each other. All links to nodes with
larger degree are made at the creation step, and then links to nodes
with smaller degree are made at each subsequent steps. This results
in the expression
\begin{eqnarray}\label{Knn1}
k_{\rm nn}(k)={1\over n_v(t_i) k(t_i,t)}\Bigg(
  \sum_{t'_i=0}^{t'_i=t_i-1} m\cdot n_v(t'_i) k(t'_i,t_i-1)k(t'_i,t) +\nonumber\\
  \qquad\qquad\qquad\qquad\qquad\sum_{t'_i=t_i+1}^{t'_i=t} m\cdot n_v(t_i) k(t_i,t'_i-1) k(t'_i,t)\Bigg)
\end{eqnarray}
for $k=2\,(m+1)^{t-t_i}$. Here the first sum on the right-hand side
accounts for the links made to nodes with larger degree (i.e.\
$t'_i<t_i$) when the node was generated at $t_i$. The second sum
describes the links made to the current smallest degree nodes at
each step $t'_i>t_i$.

Substituting Eqs. (\ref{Nv1}) and (\ref{Ki1}) into Eq. (\ref{Knn1}),
after some algebraic manipulations, Eq. (\ref{Knn1}) is simplified
to
\begin{eqnarray} \label{knn2}
k_{\rm nn}(k)= \frac{2(2m+1)}{m}\,\left [\frac{(m+1)^{2}}{2m+1}
\right]^{t_i}-\frac{2(m+1)}{m}+\frac{2m}{m+1}\,(t-t_i).
\end{eqnarray}
Thus after the initial step $k_{\rm nn}$ grows linearly with time.

Writing Eq. (\ref{knn2}) in terms of $k$, it is straightforward to
obtain
\begin{eqnarray} \label{knn3}
k_{\rm nn}(k)= \frac{2(2m+1)}{m}\,\left [\frac{(m+1)^{2}}{2m+1}
\right ]^{t}\,\left ( \frac{k}{2}\right)^{-\frac{\ln\left
[\frac{(m+1)^{2}}{2m+1}\right ]}{\ln(m+1)}}\nonumber\\
\qquad\qquad\qquad\qquad-\frac{2(m+1)}{m}+\frac{2m}{m+1}\,\frac{\ln(\frac{k}{2})}{\ln(m+1)}.
\end{eqnarray}
Therefore, $k_{\rm nn}(k)$ is approximately a power law function of
$k$ with negative exponent, which shows that the networks are
disassortative. Note that $k_{\rm nn}(k)$ of the Internet exhibit a
similar power-law dependence on the degree $k_{\rm nn}(k)\sim
k^{-\omega}$, with $\omega=0.5$ \cite{PaVaVe01}.

%\subsection{Clustering coefficient}

\emph{Clustering coefficient.} The clustering coefficient defines a
measure of the level of cohesiveness around any given node. By
definition, the clustering coefficient~\cite{WaSt98} $C_i$ of node
$i$ is the ratio between the number of edges $e_i $ that actually
exist among the $k_i $ neighbors of node $i$ and its maximum
possible value, $ k_i( k_i -1)/2 $, i.e., $ C_i =2e_i/k_i(k_i -1)$.
The clustering coefficient of the whole network is the average of
all individual $C_{i}'s$. Next we will compute the clustering
coefficient of every node and their average value.

Obviously, when a new node $i$ joins the network, its degree $k_{i}$
and $e_{i}$ is $2$ and $1$, respectively. Each subsequent addition
of a link to that node increases both $k_{i}$ and $e_{i}$ by one.
Thus, $e_{i}$ equals to $k_{i}-1$ for all nodes at all steps. So one
can see that, there is a one-to-one correspondence between the
degree of a node and its clustering. For a node with degree $k$, the
exact expression for its clustering coefficient is $2/k$. Therefore,
the clustering coefficient spectrum of nodes is discrete. Using this
discreteness, it is convenient to work with the cumulative
distribution of clustering coefficient~\cite{DoGoMe02} as
\begin{equation}
W_{\rm cum}(C)=\frac{1}{N_t}\sum_{t_j \geq
t_i}\frac{2n_v(t_j)}{k(t_j,t)}\sim
C^{{\frac{\ln(2m+1)}{\ln(m+1)}}}=C^{\gamma-1}.
\end{equation}
It is worth noting that for the special case of $m=1$, this result
has been obtained previously~\cite{DoGoMe02}.

The clustering coefficient of the whole network at arbitrary step
$t$ can be easily computed,
\begin{eqnarray}
\overline{C}_t = \frac{1}{N_{t}}\sum_{r=0}^{t}
\frac{2\,n_v(r)}{k(r,t)}
    =
    \frac{2m+2}{2m+3}\,\frac{(m+1)^{t}(2m+1)^{t}+\frac{m+2}{m+1}}{(m+1)^{t}[(2m+1)^{t}+1]}.
\end{eqnarray}
%where the sum is the total of clustering coefficient for all
%nodes and $D_r=2(m+1)^{t-r}$ shown by Eq.(~\ref{Ki}) is the
%degree of a node at step $t$ which was created at step $r$.
In the infinite network size limit ($t\rightarrow \infty$),
$\overline{C}=(2m+2)/(2m+3)$. Thus the clustering $\overline{C}$ is
high and increases with $m$. Moreover, similarly to the degree
exponent $\gamma$, $\overline{C}$ is tunable by choosing the right
value of parameter $m$: in particular, $\overline{C}$ ranges from
$4/5$ (in the special case of $m=1$~\cite{DoGoMe02}) to limit of 1
when $m$ becomes very large.

%\subsection{Diameter}

\emph{Diameter.} The diameter of a network is defined as the maximum
of the shortest distances between all pairs of nodes, which
characterizes the longest communication delay in the network. Small
diameter is consistent with the concept of small-world and it is
easy to compute for our networks. Below we give the precise
analytical computation of diameter of $\mathcal{G}(t)$ denoted by
$Diam(\mathcal{G}(t))$.

It is easy to see that at step $t = 0$ (resp. $t = 1$), the diameter
is equal to 1 (resp. 2). At each step $t\geq 2$, one can easily see
that the diameter always lies between a pair of nodes that have just
been created at this step. In order to simplify the analysis, we
first note that it is unnecessary to look at all the nodes in the
networks in order to find the diameter. In other words, some nodes
added at a given step can be ignored, because they do not increase
the diameter from the previous step. These nodes are those that
connect to edges that already existed before step $t-1$. Indeed, for
these nodes we know that a similar construction has been done in
previous steps, so we can ignore them for the computation of the
diameter. Let us call ``outer'' nodes the nodes which are connected
to a edge that did not exist at previous steps. Clearly, at each
step, the diameter depends on the distances between outer nodes.

At any step $t\geq 2$, we note that an outer node cannot be
connected with two or more nodes that were created during the same
step $t^{'}\geq t-1$. Indeed, we know that from step $2$, no outer
node is connected to two nodes of the initial triangle
$\mathcal{G}(0)$. Thus, for any step $t\geq 2$, any outer node is
connected with nodes that appeared at pairwise different steps. Now
consider two outer nodes created at step $t\geq 2$, say $v_{t}$ and
$w_{t}$. Then $v_{t}$ is connected to two nodes, and one of them
must have been created before or during step $t-2$. We repeat this
argument, and we end up with two cases: (1) $t = 2m$ is even. Then,
if we make $m$ ``jumps", from $v_{t}$ we reach the initial triangle
$\mathcal{G}(0)$, in which we can reach any $w_{t}$ by using an edge
of $\mathcal{G}(0)$ and making $m$ jumps to $w_{t}$ in a similar
way. Thus $Diam(\mathcal{G}(2m))\leq 2m+1$. (2) $t = 2m+1$ is odd.
In this case we can stop after $m$ jumps at $\mathcal{G}(1)$, for
which we know that the diameter is 2, and make $m$ jumps in a
similar way to reach $w_{t}$. Thus $Diam(\mathcal{G}(2m+1))\leq
2(m+1)$. It is easily seen that the bound can be reached by pairs of
outer nodes created at step $t$. More precisely, those two nodes
$v_{t}$ and $w_{t}$ share the property that they are connected to
two nodes that appeared respectively at steps $t-1$, $t-2$.

Hence, formally, $Diam(\mathcal{G}(t))=t+1$ for any $t\geq 0$. Note
that $N_t\sim (2m+1)^{t}$, thus the diameter is small and scales
logarithmically with the number of network nodes.

\subsection{Case of $q\geq3$}

In these cases, the analysis is a little difficult than those of the
last subsection. An alternative approach has to be adopted, although
it may also holds true for the first case in some situations. The
method of the last subsection is relatively easy to generalize to
these cases, and below we will address it, focusing on order, size,
degree distribution, clustering coefficient and diameter.

\emph{Order and size.} Let $n_v(t)$, $n_e(t)$ be the number of nodes
and edges created at step $t$, respectively. Denote $K_{q,t}$ as the
total number of $q$-cliques in the whole network at step $t$. Note
that the addition of each new node leads to $q$ new $q$-cliques and
$q$ new edges. By construction, we have $n_e(t)=qn_v(t)$,
$n_v(t)=mK_{q,t-1}$ and $K_{q,t}=K_{q,t-1}+qn_v(t)$. Thus one can
easily obtain $K_{q,t}=(mq+1)K_{q,t-1}=(q+1)(mq+1)^{t}$ ($t\geq 0$),
$n_v(t)=m(q+1)(mq+1)^{t-1}$ ($t>0$) and $n_e(t)=mq(q+1)(mq+1)^{t-1}$
($t>0$). From above results, we can easily compute the order and
size of the networks. The total number of nodes $N_t$ and edges
$E_t$ present at step $t$ is
\begin{eqnarray}\label{Nt}
N_t=\sum_{t_i=0}^{t}n_v(t_i)=\frac{(q+1)[(mq+1)^{t}+q-1]}{q}
\end{eqnarray}
and
\begin{eqnarray}\label{Et}
E_t =\sum_{t_i=0}^{t}n_e(t_i)=(q+1)(mq+1)^{t}+\frac{(q+1)(q-2)}{2},
\end{eqnarray}
respectively. For infinite $t$, the average degree $\overline{k}_t=
\frac{2E_t}{N_t}$ is approximately $2q$.

\emph{Degree distribution.} When a new node $i$ is added to the
graph at step $t_i$, it has degree $q$ and forms $q$ new
$q$-cliques. Let $n_q(i,t)$ be the total number of $q$-cliques at
step $t$ that will created new nodes connected to the node $i$ at
step $t+1$. So at step $t_i$, $n_q(i,t_i)=q$. By construction, we
can see that in the subsequent steps each new neighbor of $i$
generates $q-1$ new $q$-cliques with $i$ as one node of them. Let
$k_i(t)$ be the degree of $i$ at step $t$. It is not difficult to
find following relations for $t>t_i+1$:
\begin{equation}
\Delta k_i(t)=k_i(t)-k_i(t-1)=mn_q(i,t-1)
\end{equation}
and
\begin{equation}
n_q(i,t)=n_q(i,t-1)+(q-1)\Delta k_i(t).
\end{equation}
From the above two equations, we can derive
$n_q(i,t)=[m(q-1)+1]n_q(i,t-1)$. Considering $n_q(i,t_i)=q$, we
obtain $n_q(i,t)=q[m(q-1)+1]^{t-t_i}$ and $\Delta
k_i(t)=mq[m(q-1)+1]^{t-t_i-1}$. Then the degree $ k_i(t)$ of node
$i$ at time $t$ is
\begin{eqnarray}\label{Ki}
k_i(t)&=&k_i(t_i)+\sum_{t_h=t_i+1}^{t}{\Delta k_i(t_h)}\nonumber\\
&=&\frac{q[m(q-1)+1]^{t-t_i}+q^{2}-2q}{q-1}.
\end{eqnarray}
Since the degree of each node has been obtained explicitly as in
Eq.~(\ref{Ki}), we can get the degree distribution via its
cumulative distribution \cite{Ne03,DoGoMe02}, i.e. $P_{cum}(k)
\equiv \sum_{k^\prime \geq k} N(k^\prime,t)/N_t \sim k^{1-\gamma}$,
where $N(k^\prime,t)$ denotes the number of nodes with degree
$k^\prime$. The analytic computation details are given as follows.
For a degree $k$
\begin{equation}
k=\frac{q[m(q-1)+1]^{t-j}+q^{2}-2q}{q-1},
\end{equation}
there are  $n_v(j)=m(q+1)(qm+1)^{j-1}$ nodes with this exact degree,
all of which were born at step $j$. All nodes with birth time at $j$
or earlier have this and a higher degree. So we have
\begin{equation}
\sum_{k' \geq k} N(k',t)=\sum_{s=0}^{j}n_v(s)=\frac{(q+1)[(mq+1)^{j}+q-1]}{q}. \nonumber\\
\end{equation}
As the total number of nodes at step $t$ is given in Eq.~(\ref{Nt}),
we have
\begin{equation}
\left[\frac{q[m(q-1)+1]^{t-j}+q^{2}-2q}{q-1}\right]^{1-\gamma}\nonumber\\
=\frac{\frac{(q+1)[(mq+1)^{j}+q-1]}{q}}{\frac{(q+1)[(mq+1)^{t}+q-1]}{q}}.\nonumber\\
\end{equation}
Therefore, for large $t$ we obtain
\begin{equation}
\left [[m(q-1)+1]^{t-j}\right ]^{1-\gamma}=(mq+1)^{j-t}
\end{equation}
and
\begin{equation}\label{gamma}
\gamma \approx 1+\frac{\ln (mq+1)}{\ln[m(q-1)+1]}.
\end{equation}
For the special case $m=1$, Eq. (\ref{gamma}) recovers the results
previously reported in Ref. \cite{CoFeRa04}.

\emph{Clustering coefficient}. The analytical expression for
clustering coefficient $C(k)$ of the individual node with degree $k$
can be derived exactly. When a node is created it is connected to
all the nodes of a $q$-clique whose nodes are completely
interconnected. Its degree and clustering coefficient are $q$ and 1,
respectively. In the following steps, if its degree increases one by
a newly created node connecting to it, then there must be $q-1$
existing neighbors of it attaching to the new node at the same time.
Thus for a node of degree $k$, we have
\begin{equation}\label{Ck}
C(k)= {{{q(q-1)\over 2}+ (q-1)(k-q)} \over {k(k-1)\over 2}}=
\frac{2(q-1)(k-\frac{q}{2})}{k(k-1)},
\end{equation}
which depends on degree $k$ and $q$. For $k \gg q$, the $C(k)$ is
inversely proportional to node degree. The scaling $C(k)\sim k^{-1}$
has been found for some network
models~\cite{AlBa02,DoMe02,Ne03,BoLaMo06}, and has also observed in
several real-life networks~\cite{RaBa03}.

Using Eq. (\ref{Ck}), we can obtain the clustering $\overline{C}_t$
of the networks at step $t$:
%%%%%%%%%
%\begin{eqnarray*}
%\lefteqn{\overline{S}_t =\frac{2(q+1)(q-1)(\Delta_t-\frac{q}{2})}{\Delta_t(\Delta_t-1)}}\\
%     &+& \sum_{i=1}^{t-1} \frac{2(q+1)^{t-i}(q-1)(\Delta_i-\frac{q}{2})}{\Delta_i(\Delta_i-1)}
%\end{eqnarray*}
\begin{equation}\label{AC}
\overline{C}_t=
    \frac{1}{N_{t}}\sum_{r=0}^{t}
    \frac{2(q-1)(D_r-\frac{q}{2})n_v(r)}{D_r(D_r-1)},
\end{equation}
where the sum is the total of clustering coefficient for all nodes
and $D_r=\frac{q[m(q-1)+1]^{t-r}+q^{2}-2q}{q-1}$ shown by Eq.
(\ref{Ki}) is the degree of the nodes created at step $r$.

%%%%%%%%%%%%%%%%%%%%%%%%%%%%%%%%%%%%%%%%%%%%%%%%%%%%%%%%%%
% Figure  2
%%%%%%%%%%%%%%%%%%%%%%%%%%%%%%%%%%%%%%%%%%%%%%%%%%%%%%%%%%
\begin{figure}
\begin{center}
\includegraphics[width=9cm]{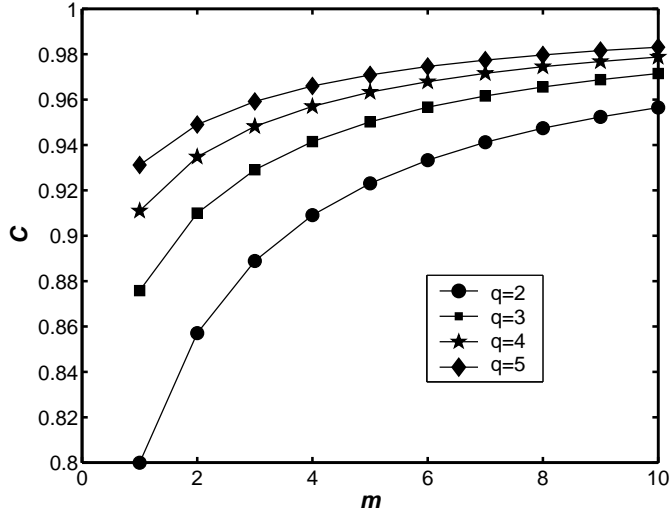}
\caption{The dependence relation of $C$ on $q$ and $m$.} \label{cc}
\end{center}
\end{figure}
%%%%%%%%%%%%%%%%%%%%%%%%%%%%%%%%%%%%%%%%%%%%%%%%%%%%%%%%%%

It can be easily proved that for arbitrary fixed $m$,
$\overline{C}_t$ increases with $q$, and that for arbitrary fixed
$q$, $\overline{C}_t$ increases with $m$. In the infinite network
order limit ($N_{t}\rightarrow \infty$), Eq. (\ref{AC}) converges to
a nonzero value $C$. When $q=2$, for $m=1$, 2, 3 and 4, $C$ equal to
0.8000, 0.8571 0.8889 and 0.9091, respectively. When $m=2$, for
$q=2$, 3, 4 and 5, $C$ are 0.8571, 0.9100, 0.9348 and 0.9490,
respectively. Therefore, the clustering coefficient of our networks
is very high. Moreover, similarly to the degree exponent $\gamma$,
clustering coefficient $C$ is determined by $q$ and $m$. Figure
\ref{cc} shows the dependence of $C$ on $q$ and $m$.

\emph{ Diameter.} In what follows, the notations $ \lceil x \rceil$
and $\lfloor x \rfloor$ express the integers obtained by rounding
$x$ to the nearest integers towards infinity and minus infinity,
respectively. Now we compute the diameter of $R(q,t)$, denoted
$Diam(R(q,t))$ for $q\geq 3$ ($q=2$ is a particular case that is
treated separately in the last subsection):

{\em Step 0}.  The diameter is $1$.

{\em Steps 1 to $\lceil\frac{q+1}{2}\rceil-1$}.  In this case, the
diameter is 2, since any new node is by construction connected to a
$q$-clique forming a $(q+1)$-clique, and since any $(q+1)$-clique
during those steps contains at least $\lceil\frac{q+1}{2}\rceil$
($q$ even) or $\lceil\frac{q+1}{2}\rceil$+1 ($q$ odd) nodes from the
initial $(q+1)$-clique $R(q,0)$ obtained after step 0. Hence, any
two newly added nodes $u$ and $v$ will be connected respectively to
sets $S_u$ and $S_v$, with $S_u\subseteq V(R(q,0))$ and
$S_v\subseteq V(R(q,0))$,  where $V(R(q,0))$ is the node set of
$R(q,0)$; however, since $\vert
S_u\vert\geq\lceil\frac{q+1}{2}\rceil$ ($q$ even) and $\vert
S_v\vert\geq\lceil\frac{q+1}{2}\rceil$+1 ($q$ odd), where $\vert
S\vert$ denotes the number of elements in set $S$, we conclude that
$S_u\cap S_v\neq{\O}$, and thus the diameter is 2.

{\em Steps $\lceil\frac{q+1}{2}\rceil$ to $q$}. In any of these
steps, some newly added nodes might not share a neighbor in the
original $(q+1)$-clique $R(q,0)$; however, any newly added node is
connected to at least one node of the initial $(q+1)$-clique
$R(q,0)$. Thus, the diameter is equal to 3.

{\em Further steps}. Similar to the case of $q=2$, we call ``outer''
nodes the nodes which are connected to a $q$-clique that did not
exist at previous steps. Clearly, at each step, the diameter depends
on the distances between outer nodes. Now, at any step $t\geq q+1$,
an outer node cannot be connected with two or more nodes that were
created during the same step $0<t'\leq t-1$. Moreover, by
construction no two nodes that were created during a given step are
neighbors, thus they cannot be part of the same $q$-clique.
Therefore, for any step $t\geq q+1$, some outer nodes are connected
with nodes that appeared at pairwise different steps. Thus, if $v_t$
denotes an outer node that was created at step $t$, then $v_t$ is
connected to nodes $v_i$s, $1\leq i\leq t-1$, where all the $i$s are
pairwise distinct. We conclude that $v_t$ is necessarily connected
to a node that was created at a step $t_0\le t-q$. If we repeat this
argument, then we obtain an upper bound on the distance from $v_t$
to the initial $(q+1)$-clique $R(q,0)$. Let $t=\alpha q+p$, where
$1\leq p\leq q$. Then, we see that $v_t$ is at distance at most
$\alpha +1$ from a node in $R(q,t)$. Hence any two
 nodes $v_t$ and $w_t$ in $R(q,0)$ lie at distance at most
$2(\alpha +1)+1$ ; however, depending on $p$, this distance can be
reduced by 1, since when $p\leq \lceil\frac{q+1}{2}\rceil-1$, we
know that two nodes created at step $p$ share at least a neighbor in
$R(q,0)$. Thus, when $1\leq p\leq \lceil\frac{q+1}{2}\rceil-1$,
$Diam(R(q,t))\leq 2(\alpha +1)$, while when
$\lceil\frac{q+1}{2}\rceil\leq p\leq q$, $Diam(R(q,t))\leq 2(\alpha
+1)+1$. One can see that these bounds can be reached by pairs of
outer nodes created at step $t$. More precisely, those two nodes
$v_t$ and $w_t$ share the property that they are connected to $q$
nodes that appeared respectively at steps $t-1,t-2,\ldots t-q$.

Based on the above arguments, one can easily see that for $t>q$, the
diameter increases by 2 every $q$ steps. More precisely, we have the
following result, for any $q\geq 3$ and $t\geq 1$ (when $t=0$, the
diameter is clearly equal to 1):
$$Diam(R(q,t))=2(\lfloor\frac{t-1}{q}\rfloor +1)+f(q,t),$$
where $f(q,t)=0$ if $t-\lfloor\frac{t-1}{q}\rfloor q\leq
\lceil\frac{q+1}{2}\rceil-1$, and 1 otherwise. When $t$ gets large,
$Diam(R(q,t))\sim \frac{2t}{q}$, while $N_t\sim (mq+1)^{t}$, thus
the diameter grows logarithmically with the number of nodes.

It is easy to see that these cases of $q\geq3$ have very similar
topological properties to the case  $q=2$. Additionally, for the
cases of $q\geq3$, the networks will again be disassortative with
respect to degree because of the lack of links between nodes with
the same degree; the second moment of degree distribution $\langle
k^{2}\rangle$ will also diverge, which is due to the fat tail of the
degree distribution.

\section{Conclusion and discussion}
To sum up, we have proposed and investigated a deterministic network
model, which is constructed in a recursive fashion. Our model is
actually a tunable generalization of the growing deterministic
scale-free networks introduced in Ref.~\cite{DoGoMe02}. Aside from
their deterministic structures, the statistical properties of the
resulting networks are equivalent with the random models that are
commonly used to generate scale-free networks
\cite{AlBa02,DoMe02,Ne03,BoLaMo06}. We have obtained the exact
results for degree distribution and clustering coefficient, as well
as the diameter, which agree well with large amount of real
observations \cite{AlBa02,DoMe02,Ne03,BoLaMo06}. The degree exponent
can be adjusted, the clustering coefficient is very large, and the
diameter is small. Therefore, out model may perform well in
mimicking a variety of scale-free networks in real-life world.
Moreover, our networks consist of cliques, which has been observed
in variety of the real-world networks, such as movie actor
collaboration networks, scientific collaboration networks and
networks of company directors \cite{AlBa02,DoMe02,Ne03,BoLaMo06}.
\smallskip

\subsection*{Acknowledgment}
This research was supported in part by the National Natural Science
Foundation of China (NNSFC) under Grant Nos. 60373019, 60573183, and
90612007. Lili Rong gratefully acknowledges partial support from
NNSFC under Grant Nos. 70431001 and 70571011. The authors thank the
anonymous referees for their valuable comments and suggestions.

%%%%%%%%%%%%%%%%%%%%%%%%%%%%%%%%%%%%%%%%%%%%%%%%%%%%%%%%%%%%%%%%%
%%%%%%%%%%%%%%%%%%%%%%%%%%%%%%%%%%%%%%%%%%%%%%%%%%%%%%%%%%%%%%%%%


\begin{thebibliography}{10}
%% 1
\bibitem{WaSt98} D. J. Watts and H. Strogatz,
       %Collective dynamics of `small-world' networks,
        Nature (London) {\bf 393}, 440 (1998).


%% 2
\bibitem{BaAl99} A.-L. Barab\'asi and R. Albert,
      %Emergence of scaling in random networks,
       Science {\bf 286}, 509 (1999).

%% 3
\bibitem{St01}
S. H. Strogatz, Nature {\bf 410}, 268 (2001).

%% 4
\bibitem{AlBa02} R. Albert and A.-L. Barab\'asi,
      %Statistical mechanics of complex networks,
       Rev. Mod. Phys. {\bf 74}, 47 (2002).

%% 5
\bibitem{DoMe02} S. N. Dorogvtsev and J.F.F. Mendes,
%Evolution of networks,
Adv. Phys. {\bf 51}, 1079 (2002).

%% 6
\bibitem{Ne03} M. E. J. Newman,
%The structure and function of complex networks,
SIAM Review {\bf 45}, 167 (2003).


%% 8
\bibitem{BoLaMo06}
S Boccaletti, V Latora, Y Moreno, M. Chavezf, and D.-U. Hwanga,
 %Complex networks: Structure and dynamics.
Physics Report {\bf 424}, 175 (2006).

%% 9
\bibitem{BaRaVi01} A.-L. Barab\'asi, E. Ravasz, and T. Vicsek,
          %Deterministic scale-free networks,
          Physica A  {\bf 299}, 559 (2001).

%% 10
\bibitem{IgYa05}
K. Iguchi and H. Yamada, Phys. Rev. E {\bf 71}, 036144 (2005).


%% 11
\bibitem{DoGoMe02} S. N. Dorogovtsev, A. V. Goltsev, and J. F. F. Mendes,
           %Pseudofractal scale-free web,
          Phys. Rev. E {\bf 65}, 066122 (2002).

%% 12
\bibitem{CoFeRa04} F. Comellas, G. Fertin and A. Raspaud,
%Recursive graphs with small-world scale-free properties,
Phys. Rev. E {\bf 69}, 037104 (2004).


%% 13
\bibitem{JuKiKa02} S. Jung, S. Kim, and B. Kahng,
       %Geometric fractal growth model for scale-free networks.
        Phys. Rev. E {\bf 65}, 056101 (2002).



%% 14
\bibitem{RaSoMoOlBa02}
E. Ravasz, A.L. Somera, D.A. Mongru, Z.N. Oltvai, and A.-L.
Barab\'asi, Science {\bf 297}, 1551 (2002).


%% 15
\bibitem{RaBa03}
E. Ravasz and A.-L. Barab\'asi, Phys. Rev. E {\bf 67}, 026112
(2003).

%% 16
\bibitem{No03}
J. D. Noh, Phys. Rev. E {\bf 67}, 045103 (2003).

%% 17
\bibitem{NaUeKaAk05}
J. C. Nacher, N. Ueda, M. Kanehisa and T. Akutsu, Phys. Rev. E {\bf
71}, 036132 (2005).


%% 18
\bibitem{AnHeAnSi05} J. S. Andrade Jr., H. J. Herrmann, R. F. S. Andrade and L. R. da Silva,
%Apollonian Networks: Simultaneously scale-free, small world, Euclidean, space filling,
%and with matching graphs.
Phys. Rev. Lett. {\bf 94}, 018702 (2005).

%% 19
\bibitem{DoMa05} J. P. K. Doye and C. P. Massen,
%Self-similar disk packings as model spatial scale-free networks.
Phys. Rev. E {\bf 71}, 016128 (2005).

%% 20
\bibitem{RoKiBobe05}H. Rozenfeld, J. Kirk, E. Bollt and D. ben-Avraham, J. Phys. A {\bf 38},
4589 (2005).

%% 21
\bibitem{Bobe05}E. Bollt, D. ben-Avraham, New Journal of Physics {\bf 7}, 26
(2005).

%% 22
\bibitem{HiBe06}
M. Hinczewski and A. N. Berker, Phys. Rev. E {\bf 73}, 066126
(2006).

%% 23
\bibitem{ZhCoFeRo05}
Z. Z. Zhang, F. Comellas, G. Fertin and L. L. Rong,
%High dimensional Apollonian networks.
 J. Phys. A {\bf 39}, 1811 (2006).


%% 24
\bibitem{ZhRo05} Z. Z. Zhang, L. L. Rong, and Shuigeng Zhou,
 Phys. Rev. E, {\bf 74}, 046105
(2006).


% 25
\bibitem{CoOzPe00}
F. Comellas, J. Oz\'on, and J. G. Peters, Inf. Process. Lett., {\bf
76}, 83 (2000)

%26
\bibitem{CoSa02}
F. Comellas and M. Sampels, Physica A {\bf 309}, 231 (2002).


%27
\bibitem{ZhRoGo05}
Z. Z. Zhang, L. L. Rong and C. H. Guo, Physica A {\bf 363}, 567
(2006).

%28
\bibitem{ZhRoCo05a} Z. Z. Zhang, L. L. Rong and F. Comellas,
%High dimensional random Apollonian networks.
J. Phys. A {\bf 39}, 3253 (2006).

%29
\bibitem{ZhWaHuCh04}
T. Zhou, B. H. Wang, P. M. Hui and K. P. Chan, Physica A {\bf 367},
613 (2006).

%30
\bibitem{co04}
G. Corso, Phys. Rev. E {\bf 69}, 036106 (2004).

%31
\bibitem{Ac04}
J. D. Achter, Phys. Rev. E {\bf 70}, 058103 (2004).

%32
\bibitem{ChDa05}
A. K. Chandra, S Dasgupta,
%A small world network of prime numbers.
Physica A, Physica A {\bf 357}, 436 (2005).


%% 33
\bibitem{ZhYaWa05} T. Zhou, G. Yan, and B. H. Wang,
%Maximal planar networks with large clustering coefficient and power-law degree distribution
Phys. Rev. E {\bf 71}, 046141 (2005).

%% 34
\bibitem{ZhRoCo05b} Z. Z. Zhang, L. L. Rong and F. Comellas,
%High dimensional random Apollonian networks.
Physica A {\bf 364}, 610 (2006).


%% 35
\bibitem{AnHe05}
R. F. S. Andrade and H. J. Herrmann,
%Magnetic models on Apollonian networks
Phys. Rev. E {\bf 71}, 056131 (2005).

%% 36
\bibitem{LiGaHe04} P. G. Lind, J.A.C. Gallas, and H.J. Herrmann,
%Magnetic models on Apollonian networks
Phys. Rev. E {\bf 70}, 056207 (2004).

%% 33
\bibitem{AnMi05} R.F.S. Andrade, J.G.V. Miranda,
%Spectral properties of the Apollonian network
Physica A {\bf 356}, 1 (2005).


%% 33
\bibitem{GrKa82}
R. B. Griffiths and M. Kaufman, Phys. Rev. B {\bf 26}, 5022 (1982).

%% 40
\bibitem{PaVe01a} R. Pastor-Satorras and A. Vespignani, Phys. Rev. Lett. {\bf 86}, 3200
(2001).

%% 40
\bibitem{PaVe01b}
R. Pastor-Satorras and A. Vespignani, Phys. Rev. E {\bf 63}, 066117
(2001).

%% 40
\bibitem{AlJeBa00}R. Albert, H. Jeong, and A.-L. Barab\'asi, Nature (London) {\bf 406},
378 (2000).

%% 40
\bibitem{CoerAvSa01}
R. Cohen, K. Erez, D. ben Avraham, and S. Havlin, Phys. Rev. Lett.
{\bf 86}, 3682 (2001).

\bibitem{MsSn02}
S. Maslov and K. Sneppen,
%Specificity and stability in topology of
%protein  networks,
Science {\bf 296}, 910 (2002).

\bibitem{PaVaVe01}
R. Pastor-Satorras, A. V\'azquez and A. Vespignani,
% Dynamical and correlation properties of the {I}nternet,
Phys. Rev. Lett. {\bf 87}, 258701 (2001).

\bibitem{VapaVe02}
A. V\'azquez, R. Pastor-Satorras and A. Vespignani,
%Large-scale topological  and dynamical properties of the {I}nternet,
Phys. Rev. E {\bf 65}, 066130 (2002).

\bibitem{Newman02}
M. E.~J. Newman,
%Assortative mixing in networks,
Phys. Rev. Lett. {\bf 89}, 208701 (2002).

\bibitem{Newman03c}
M. E.~J. Newman,
% Mixing patterns in networks,
Phys. Rev. E {\bf 67}, 026126 (2003).


\end{thebibliography}
\end{document}